\documentclass[twocolumn,showpacs,a4paper]{revtex4} 
\usepackage{epsfig}
\usepackage{amsbsy}

\begin{document}

\title{Competition--induced shifts of spectral peaks 
in photon coincidence spectroscopy}

\author{Levente Horvath
	and Barry C.\ Sanders}

\affiliation{Department of Physics, Macquarie University, 
Sydney, New South Wales 2109, Australia}

\date{Received: 2000 / Revised version: } 

\begin{abstract}
We show that shifts in locations of two--photon coincidence 
spectral peaks, 
for a bichromatically--driven two--level atom passing through a single--mode
cavity, are due to competition between excitation pathways for
a Jaynes--Cummings system. We also discuss an analogous shift of
(single--photon) spectral peaks for a driven three--level $\vee$--system,
which demonstrates that competition between excitation pathways is also
important in this simple system.
\end{abstract}
\pacs{42.50.Ct, 42.50.Dv}

\maketitle 

\section{Introduction}
\label{intro}

Cavity quantum electrodynamics (CQED) has continued to develop rapidly, 
driven both by recent experimental successes and by the promise 
of exciting new applications. Advances in atom cooling techniques, as well
as development of high--Q optical cavities with large--dipole coupling, 
have enabled testing of the strong--coupling regime of 
CQED~\cite{Tho98}
as well as the trapping of
single atoms in optical cavities~\cite{Hoo00,Pin00}. 

Applications of CQED, especially for such applications as the quantum
logic gate~\cite{Tur95}, rely critically on
entanglement between the field degree of freedom and the internal
electronic state of the atom~\cite{Bru96,Car96}. 
This entanglement is not only challenging to achieve,
it is also difficult to probe.
In the optical regime of CQED, photon coincidence
spectroscopy (PCS) has been proposed as a feasible and unambiguous method
for detecting genuine quantum phenomena in CQED. This technique
employs a bichromatic (or
multichromatic) driving field acting on the
combined atom--cavity system and detects two--photon (or 
multiphoton) decays, respectively, by registering
photon coincidences in the cavity field emission~\cite{Car96,San97,Hor99}.

The simplest case of PCS is two--photon coincidence spectroscopy (2PCS)
for probing the nonlinear portion of the Jaynes--Cummings (JC) spectrum. 
The technique of 2PCS proceeds,
first by driving the atomic beam with a bichromatic field
(consisting of a fixed driving field with fixed frequency~$\omega_1$ and
a scanning field with tunable frequency~$\omega_2$), which
causes two--photon excitation to the second couplet of the JC ladder,
followed by two--photon decay from the atom--cavity system 
(Fig.~\ref{fig:ladder}).
The objective is to count photon pairs emitted from the cavity
as a function of varying~$\omega_2$.
When the sum frequency for the bichromatic driving field~$\omega_1+\omega_2$
matches a 
transition frequency from the second couplet in the JC
spectrum to the ground state, the result is
an enhanced two--photon count rate (2PCR). 

This simple picture of excitation pathways agrees with 
simulations~\cite{Car96,San97}, but
displacement of coincidence peaks as
a function of scanning field frequency was evident yet not properly
understood~\cite{Car96}. Here we show that the deviation of peak shifts
is due to competition between excitation pathways. This picture yields
excellent quantitative agreement with simulations of 2PCS.

\section{Master equation and two--photon count rates}
\label{sec:Master:Twophoton}

In the electric--dipole and
rotating--wave approximations, the JC Hamiltonian for the two--level atom 
(2LA) coupled with
the single mode is
\begin{equation} 
\label{JCH} 
H(g) = \omega (\sigma_z + a^{\dagger} a)
	+ i g ( \mbox{\boldmath{$r$}} ) (a^{\dagger} \sigma_- - a \sigma_+ ), 
\end{equation}
with~$\mbox{\boldmath{$r$}}$ the position of the atom,
$g(\mbox{\boldmath{$r$}})$ the position--dependent dipole coupling strength, 
$a$ and $a^{\dagger}$ the annihilation and creation operators 
for photons in the cavity field, 
$\sigma_+$, $\sigma_-$, and $\sigma_z$ the 2LA  
raising, lowering and inversion operators, respectively, 
and $\hbar = 1$. 
Provided that the atoms move sufficiently slowly through the cavity 
\cite{Car96,San97},
the atom can be treated as if it were
at rest at some randomly located position~$\mbox{\boldmath{$r$}}$.
As the position~$\mbox{\boldmath{$r$}}$ is a randomly varying quantity,   
the value of the coupling strength~$g$ itself is also random. 
Hence, a coupling strength distribution~$P(g)$ can be  
constructed~\cite{San97}, and we assume the~$P(g)$ depicted in 
Fig.~5 of Ref.~\cite{San97}. This~$P(g)$ corresponds to the coupling
strength distribution obtained for atoms passing through a 
rectangular mask, centered at an antinode, for the cavity sustaining only 
the~TEM$_{00}$ mode. 
The dimension of the mask is~$w_0$ (cavity mode waist) by~$\lambda/10$ 
(with~$\lambda$ the optical wavelength). 
We restrict~$Fg_{\rm max}<g<g_{\rm max}$ for~$g_{\rm max}$ the 
coupling strength at an antinode along the cavity longitudinal
axis and~$F$ an effective cut--off term.

\begin{figure}
\begin{picture}(200,300)(0,-10)
\font\gnuplot=cmr10 at 10pt
\gnuplot
{\resizebox{220pt}{270pt}{\includegraphics{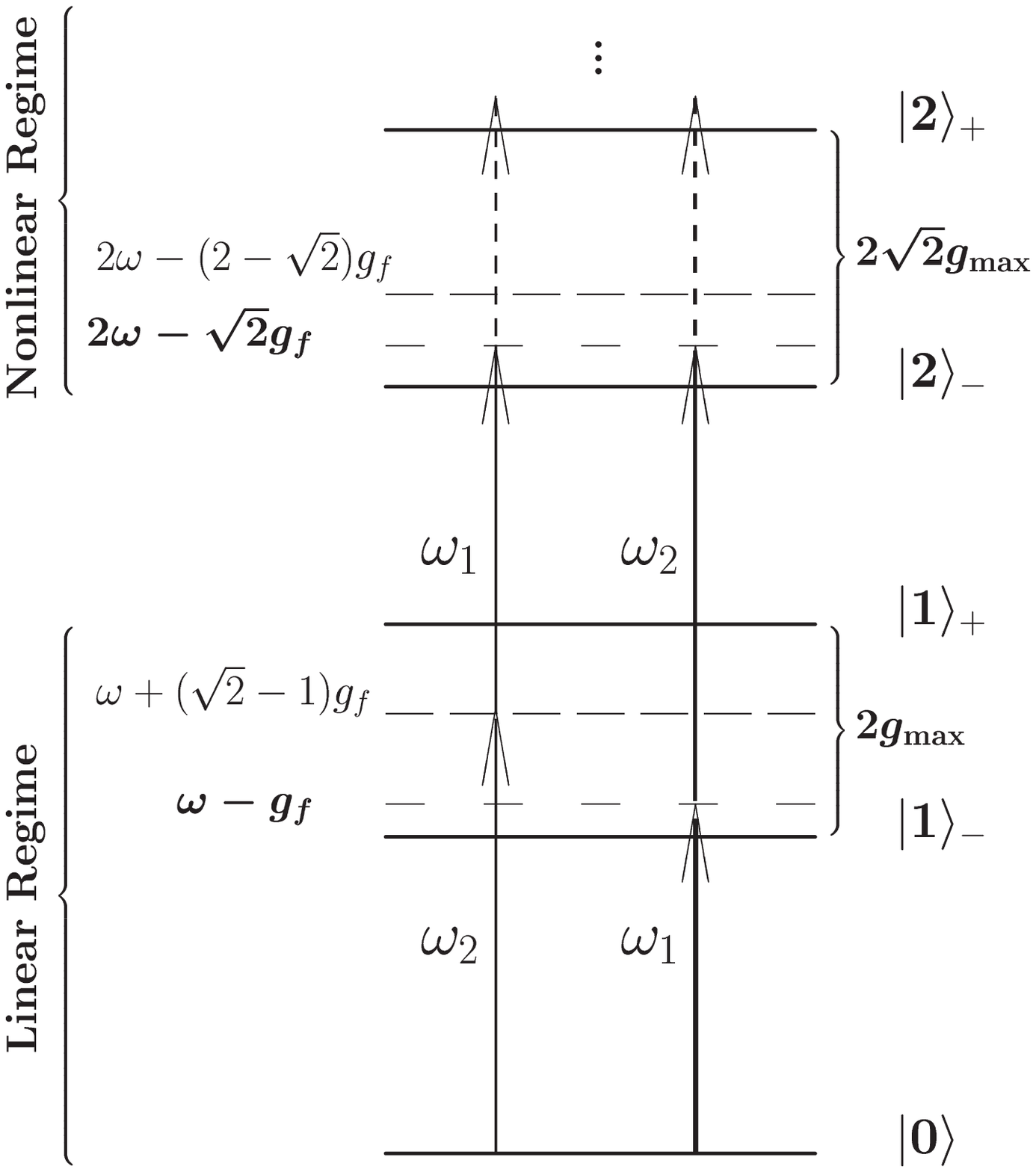}}}
\end{picture}
\caption{Two--photon excitation scheme from the ground state 
	$\left\vert 0 \right\rangle$ to the first two excited couplets
	$\left\vert n \right\rangle_{\varepsilon}$ 
	($n\in\{1,2\}$, $\varepsilon \in \{-,+\}$) of the dressed states.
	The inhomogeneous broadening of the 
	couplets~$\left\vert 1 \right\rangle_\varepsilon$ 
	and~$\left\vert 2 \right\rangle_\varepsilon$ 
	is~$2\hbar g_{\rm max}$ and~$2\sqrt2 \hbar g_{\rm max}$, respectively. 
	Two two--photon excitations to the second couplet are
	depicted for a bichromatic driving field with one component
	of amplitude~${\cal E}_1$ and the other with amplitude~${\cal E}_2$.
	The excitation pathway on the right ($\omega_1$ then~$\omega_2$)
	excites resonantly from~$\vert 0\rangle$ to~$\vert 1\rangle_-$
	and then may excite resonantly to either~$\vert 2\rangle_{\pm}$.
	The excitation pathway on the left ($\omega_2$ then~$\omega_1$)
	excites resonantly from~$\vert 0\rangle$ to~$\vert 1\rangle_+$
	to~$\vert 2\rangle_-$ for~$g=(\sqrt2-1)g_f$.}
\label{fig:ladder}
\end{figure}

The spectrum for the Hamiltonian (\ref{JCH}), 
depicted in Fig.~\ref{fig:ladder},
is the well--known JC spectrum, or `ladder'~\cite{Jay63}.
The `dressed states' of the combined atom--cavity system
are here designated by the lowest--energy 
state~$|0\rangle \equiv \vert 0 \rangle_{\rm cav}
	\otimes \vert {\tt g} \rangle_{\rm atom} 
	\equiv \vert 0, {\tt g} \rangle$,
and, for $n$ a positive integer, the `excited' 
states~$|n \rangle_{\pm} \equiv i/\sqrt{2}
	 \left( \left\vert n-1, {\tt e} \right\rangle
	 \pm i \left\vert n, {\tt g} \right\rangle \right)$,
with~$ |n\rangle $ the Fock state of the cavity mode and
$ |{\tt g}\rangle \, ( |{\tt e}\rangle )$
the ground (excited) state of the 2LA. 
The effect of averaging over~$P(g)$ is an inhomogeneous spectral 
broadening, due to atomic position variability,
as shown in~Fig.~\ref{fig:ladder}.

Two--photon excitation is provided by driving the atom directly, as it
traverses the cavity, 
with a bichromatic field~${\cal E}(t) = {\cal E}_1 e^{-i\omega_1 t} 
	+ {\cal E}_2 e^{-i\omega_2 t}$.
The driving--field frequency~$\omega_1$ is fixed and resonantly excites the
atom--cavity system from the ground state~$|0\rangle$ to the
excited state~$|1\rangle_-$  for the subensemble~$g=g_f=\omega-\omega_1$;
this subensemble corresponds to~$P(g)=\delta(g-g_f)$.

The scanning--field frequency is $\omega_2$,
 and excites the subensemble for~$g=g_f$ from~$\vert 1\rangle_-$
to one of the two states
in the second couplet of the JC ladder,
namely $|2\rangle_{\pm}$.
Thus, the range of scanning frequencies for $\omega_2$ must
include the $|1\rangle_- \longleftrightarrow |2\rangle_{\pm}$
transition frequencies,
$\omega \pm ( \sqrt{2} \mp 1 ) g$,
respectively.
The amplitudes of the two chromatic components should be large enough 
to ensure sufficient occupation of the excited state but not large 
enough that significant Stark shifting or nonnegligible occupation of the 
higher--order states occurs.
Enhanced rates of photon pair detection are then sought as the
scanning frequency~$\omega_2$ is varied 
such that~$\omega_1+\omega_2$ is resonant
with some transition~$\vert 0\rangle\longleftrightarrow \vert 2\rangle_{\pm}$
as depicted in Fig.~\ref{fig:ladder}.

The Born--Markov approximation is applied to both radiation reservoirs:
the reservoir for the field leaving the cavity and the reservoir
for direct fluorescence of the 2LA from the sides of the cavity
into free space.
The cavity damping rate is~$\kappa$,
and the 
cavity--inhibited spontaneous
emission rate into free space is~$\gamma$.
The master equation for this system~\cite{San97} can be expressed as
$\dot{\rho} = {\cal L}\rho$ for~$\cal L$ the Liouville operator.
Here ${\cal L}={\cal L}_{\rm eff}+{\cal D}+{\cal J}$, i.e.~a 
sum of a Liouville operator~${\cal L}_{\rm eff}$,
an explicit time--dependent Liouville operator ${\cal D}$ and a `jump' 
term~${\cal J}$.
We introduce $\delta \equiv \omega_2-\omega_1$
and work in the rotating picture with respect to the driving--field
component~$\omega_1$.

The effective Hamiltonian is (without jump terms)
\begin{eqnarray}
\label{eq:Heff}
H_{\rm eff}(g,{\cal E}_1)
	&=& \left( \omega - \omega_1 \right) (\sigma_z + a^{\dagger} a) 
	+ \Xi(g)+\Upsilon({\cal E}_1)	\nonumber	\\
	&&- i\kappa a^{\dagger} a - i(\gamma/2) \sigma_+ \sigma_-,
\end{eqnarray}
for~$\Xi(g)=i g (a^{\dagger} \sigma_- - a \sigma_+ )$ the
quantum exchange operator and
$\Upsilon({\cal E}_1)=i{\cal E}_1 (\sigma_+-\sigma_-)$ a monochromatic 2LA
driving
term. The corresponding Liouville operator is 
\begin{equation}
\label{eq:Leff}
{\cal L}_{\rm eff} (g, {\cal E}_1) \rho
	= -i \left[ H_{\rm eff}(g, {\cal E}_1) \rho 
		- \rho H_{\rm eff}^{\dagger} (g, {\cal E}_1) \right] .
\end{equation}
The operator for the jump term is~$\cal J$ such that
\begin{equation}
\label{Liouvillean:J}
{\cal J} \rho = 2 \kappa a \rho a^{\dagger}+\gamma \sigma_- \rho \sigma_+,
\end{equation}
and the time--dependent Liouville operator is ${\cal D}(t)$ such that
\begin{equation}
\label{Liouvillean:D}
{\cal D} \rho
	= -i \left[ \Upsilon({\cal E}_2 e^{-i\delta t}),\rho \right].
\end{equation}
Solving the master equation for $\dot{\rho}$, and averaging over $P(g)$,
which accounts for atomic position variability, yields a solution
\begin{equation}
\label{density:matrix}
\bar{\rho} \equiv \int_{Fg_{\rm max}}^{g_{\rm max}} P(g) \rho(g) dg. 
\end{equation}

For a bichromatic driving field, the density 
matrix does not settle to a steady state value, but 
in the long--time limit~$t \longrightarrow \infty$, the
Bloch function expansion is
\begin{equation}
\label{Bloch}
\lim_{t\rightarrow \infty} 
\bar{\rho}(t) = \sum_{m=-\infty}^{\infty} \bar{\rho}_m e^{i m \delta t},
\end{equation}
with
$\overline{\rho}_m$
time--independent matrices.
As the photocount integration time is expected to be long compared to the 
detuning~$\delta$, it is reasonable to assume that rapidly oscillating
terms average out and therefore
approximate $\rho(g)$
by truncating the expansion~(\ref{Bloch}). 

\begin{figure}
\begin{picture}(200,180)(60,10)
\font\gnuplot=cmr10 at 10pt
\gnuplot
\rotatebox{90}{\resizebox{250pt}{330pt}{\includegraphics{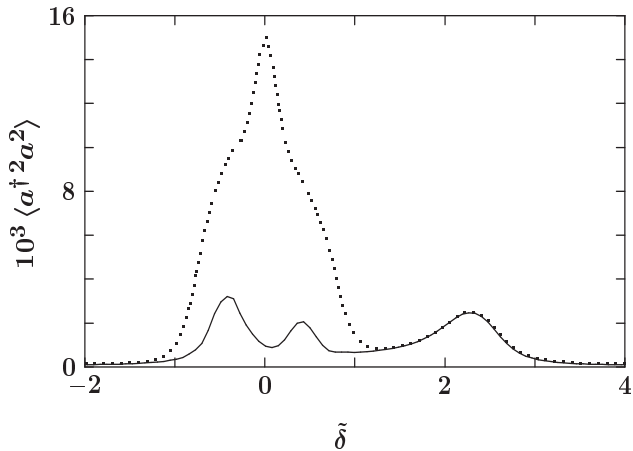}}}
\end{picture}
\caption{The 2PCR vs the normalized detuning~$\tilde\delta$ for the
inhomogeneously broadened system, with  
${\cal E}_1/\kappa=1/\sqrt2$, ${\cal E}_2/\kappa=\sqrt2$, $g_f/\kappa=9$, 
$\gamma/\kappa=2$: $w^{(2)}(\delta,{\cal E}_1)$ is the solid line
and~$\Delta^{(2)}(\delta,{\cal E}_1)$ is the dotted line.
}
\label{fig:Fig1}
\end{figure}

The optical signature for entanglement is the 2PCR~\cite{Hor99},
\begin{equation}
\label{eq:w2}
w^{(2)}(\delta,{\cal E}_1)
	= \left\langle : n^2 : \right\rangle (\delta,{\cal E}_1),
\end{equation} 
for $n \equiv a^{\dagger} a$. 
The 2PCR is shown in~Fig.~\ref{fig:Fig1} as a dotted line, and the
peaks expected at~$\tilde\delta=(\delta-g_f)/g_f=\pm(\sqrt2-1)$
are not distinguishable. This indistinguishability is due to 
off--resonant two--photon transitions with both photons of
frequency~$\omega_2$, and
may be overcome via the method of background subtraction:
the experiment is repeated with~${\cal E}_1=0$ and the resultant 2PCR
subtracted. Thus, we define a difference--2PCR
\begin{equation}
\Delta^{(2)}(\delta, {\cal E}_1) = w^{(2)}(\delta,{\cal E}_1)
	- w^{(2)}(\delta,{\cal E}_1=0).
\end{equation}

For a specific set of parameters
we show in Fig.\ {\ref{fig:Fig1}} the 2PCR with and 
without background
subtraction, which is significant in the domain of homogeneous 
broadening~$-1\leq \tilde\delta\leq 1$,
but the region near~$\tilde\delta=1+\sqrt2$
is not significantly 
affected by background subtraction. Background subtraction 
helps to resolve 
the two two-photon spectral peaks at~$\tilde\delta=\pm(\sqrt2-1)$.

\section{Competition--induced spectral peak shifts}
\label{sec:CISPS}

\begin{figure}
\begin{picture}(200,180)(60,10)
\font\gnuplot=cmr10 at 10pt
\gnuplot
\rotatebox{90}{\resizebox{250pt}{330pt}{\includegraphics{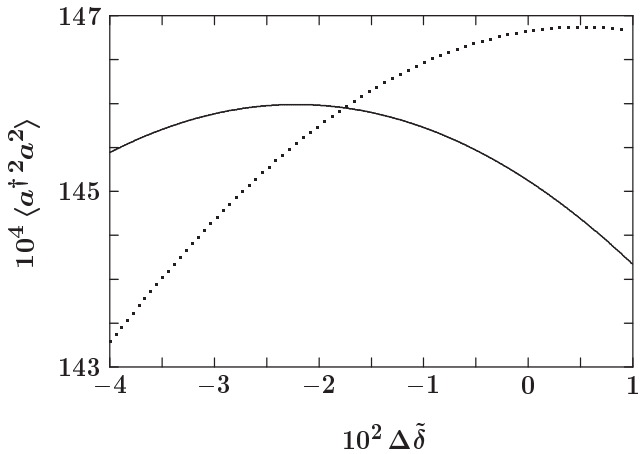}}}
\end{picture}
\caption{Two--photon count rate (2PCR) vs the shift in normalized 
	scanning 
	frequency of the peak at~$\tilde\delta=\sqrt{2}+1$ 
	for~$g/\kappa=9$, ${\cal E}_1/\kappa=1/\sqrt{2}$,
	${\cal E}_2/\kappa=\sqrt{2}$ and $\gamma/\kappa=2$ depicted as
	the solid line. 
	The dotted line corresponds to the 
	case with condition~(\ref{pathwayrestriction}) imposed.}
\label{fig:shift}
\end{figure}

The objective of 2PCS is to obtain
a 2PCR with peaks at values of~$\tilde\delta$ which cannot be explained
by semiclassical models. The peaks at~$\tilde\delta=\pm(\sqrt2-1)$ 
and~$\tilde\delta=\sqrt2+1$ involve the quantity~$\sqrt2$, which follows
from the quantum field expression
$a\vert 2\rangle=\sqrt2\vert 1 \rangle$. This~$\sqrt2$ is 
therefore a genuine, 
unambiguous quantum field signature. However, the actual peaks in 2PCR 
cannot be expected to occur at exactly these values of~$\tilde\delta$ as
observed quite clearly in Fig.~5 of Ref.\ \cite{Car96}.
In Fig.~\ref{fig:shift} we show in detail the peak near~$\tilde\delta
=\sqrt2+1$ as a solid line
for~$P(g)=\delta(g-g_f)$
and an anomalous shift from its expected position at~$\tilde\delta=\sqrt2+1$,
or~$\Delta\tilde\delta\equiv\tilde\delta-(\sqrt2+1)=0$, is evident. 
We begin exploring this anomalous shift for the~$g=g_f$
subensemble.

We observe that this peak is noticeably shifted from the
expected location.
The origin of the shifts in peaks is subtle. In Ref.\ \cite{Car96} the 
suggestion is that ``the peaks are shifted slightly, probably due to 
the motion of atoms''. Although atomic motion may contribute to peak
shifts, we show in this section that competition between excitation
pathways is the major factor in shifts of 2PCS peaks. 

The peak at~$\tilde\delta=\sqrt2+1$ is the most interesting
 of the three major 2PCR peaks (i) because
background subtraction is not required to detect this peak 
(as~$\tilde\delta$ lies outside the region of inhomogeneous broadening)
and (ii) because the peak is quite distinct from the large neighboring pair of
peaks at~$\tilde\delta=\pm(\sqrt2-1)$.
The excitation pathway to obtain the peak at~$\tilde\delta=\sqrt2+1$
is via an~$\omega_1$ photon which resonantly excites~$\vert 0\rangle
\longleftrightarrow \vert 1 \rangle_-$, followed by an~$\omega_2$
photon which resonantly excites~$\vert 1\rangle_-
\longleftrightarrow \vert 2 \rangle_+$ for 
the 2PCR peak at~$\tilde\delta=\sqrt2+1$ as shown in Fig.~\ref{fig:ladder}. 
However, level~$\vert 2\rangle_-$ becomes populated as well 
as~$\vert 2\rangle_+$ by off--resonant excitation.

The competition between populating 
levels~$\vert 2\rangle_{\pm}$ 
is evident in Figs.\ 3(d,e) of Ref.\ \cite{San97}, 
which depict the populations of levels~$\vert 2\rangle_{\pm}$ 
as functions of~$g$ 
and~$\tilde\delta$. The~$\vert 2\rangle_-$ level is noticeably occupied
for all values of~$\tilde\delta$ for~$g$ in the vicinity of~$g_f$.
This occupation of~$\vert 2\rangle_-$ was not explicitly discussed in
Ref.\ \cite{Car96,San97}, and the significance of this population 
of~$\vert 2\rangle_-$ was not provided with respect to the shift in
the peak at~$\tilde\delta=\sqrt2+1$. The shift in the peak 
is shown but not
discussed in Ref.\ \cite{San97}. The following analysis provides the first
detailed picture for explaining the shift in the 2PCR peaks.

The effect of competition is made clear in Fig.~\ref{fig:shift}. Here
we present one portion of the 2PCR peak as a function of~$\tilde\delta$,
in the vicinity of~$\Delta\tilde\delta=0$. 
The solid line corresponds
to the 2PCS. The dotted line corresponds to the simulated 2PCS
but with the artificial restriction that the matrix elements of the
Hamiltonian driving field 
operator~$\Upsilon({\cal E}_2 \exp(-i\delta t))$
for the transition~$\vert 1 \rangle_-
\longleftrightarrow \vert 2\rangle_-$ are set to zero. That is,
we artificially impose the condition 
\begin{equation}
\label{pathwayrestriction}
_-\langle 1\vert \Upsilon ({\cal E}_2 \exp(-i\delta t)) 
\vert 2 \rangle_-
	=0
	=_-\!\!\langle 2\vert\Upsilon({\cal E}_2\exp(-i\delta t))
	\vert 1\rangle_-~.
\end{equation}
By doing
so we observe, and show in Fig.~\ref{fig:shift}, that the 
modified 2PCS graph is indeed centered at~$\Delta\tilde\delta=0$
(equivalently, at~$\tilde\delta=\sqrt2+1$). The comparison between the
two cases,
with both~$_-\langle 1\vert\Upsilon({\cal E}_2 \exp(-i\delta t)) 
\vert 2 \rangle_-$ and
its conjugate set to zero vs both not being set to zero, provides
convincing evidence that the shift in the peak is indeed
primarily due to competition between two excitation pathways. 

\begin{figure}
\begin{picture}(200,280)(63,60)
\font\gnuplot=cmr10 at 10pt
\gnuplot
\rotatebox{90}{\resizebox{250pt}{330pt}{\includegraphics{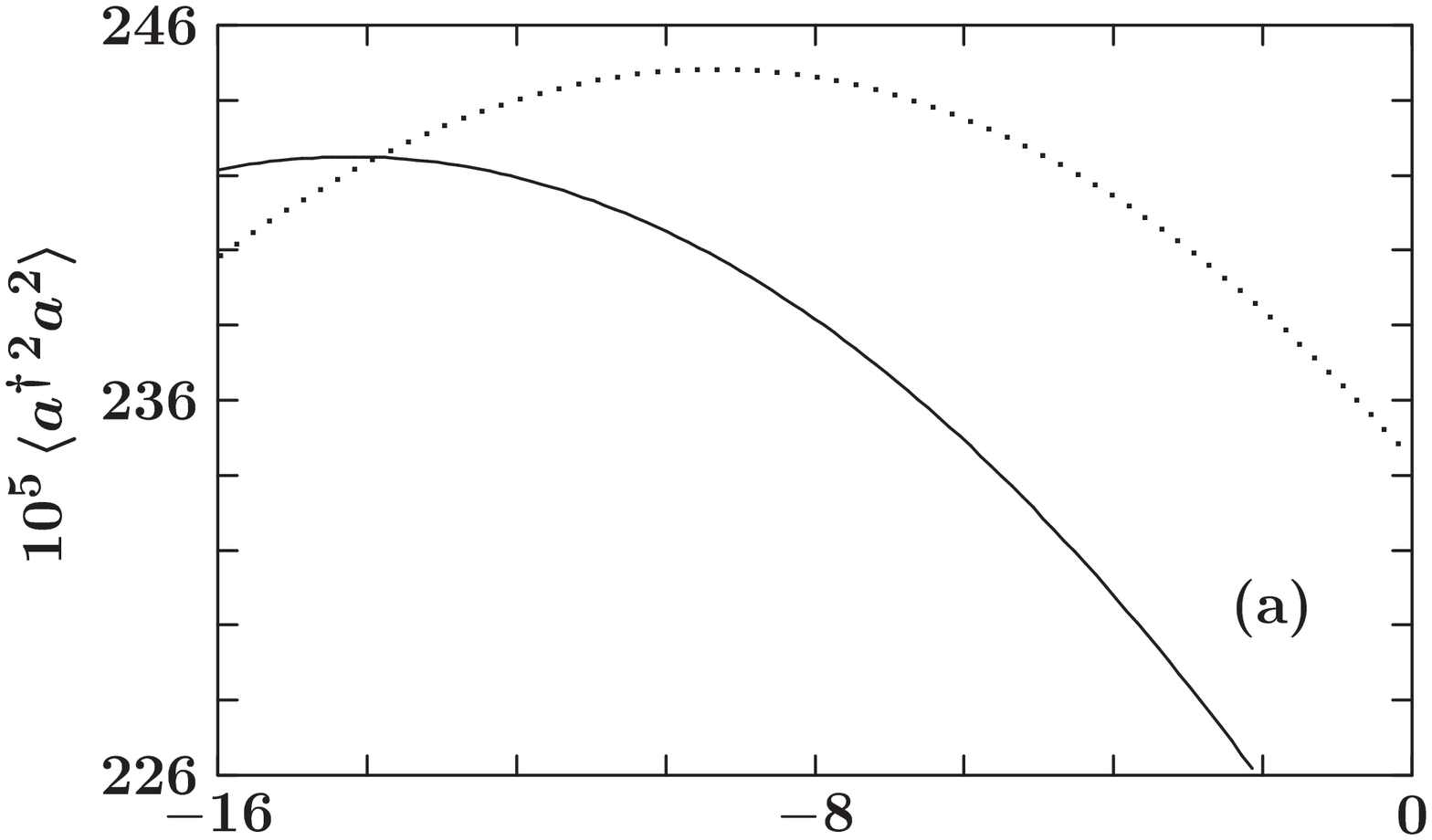}}}
\end{picture}
\begin{picture}(200,220)(60,0)
\font\gnuplot=cmr10 at 10pt
\gnuplot
\rotatebox{90}{\resizebox{250pt}{330pt}{\includegraphics{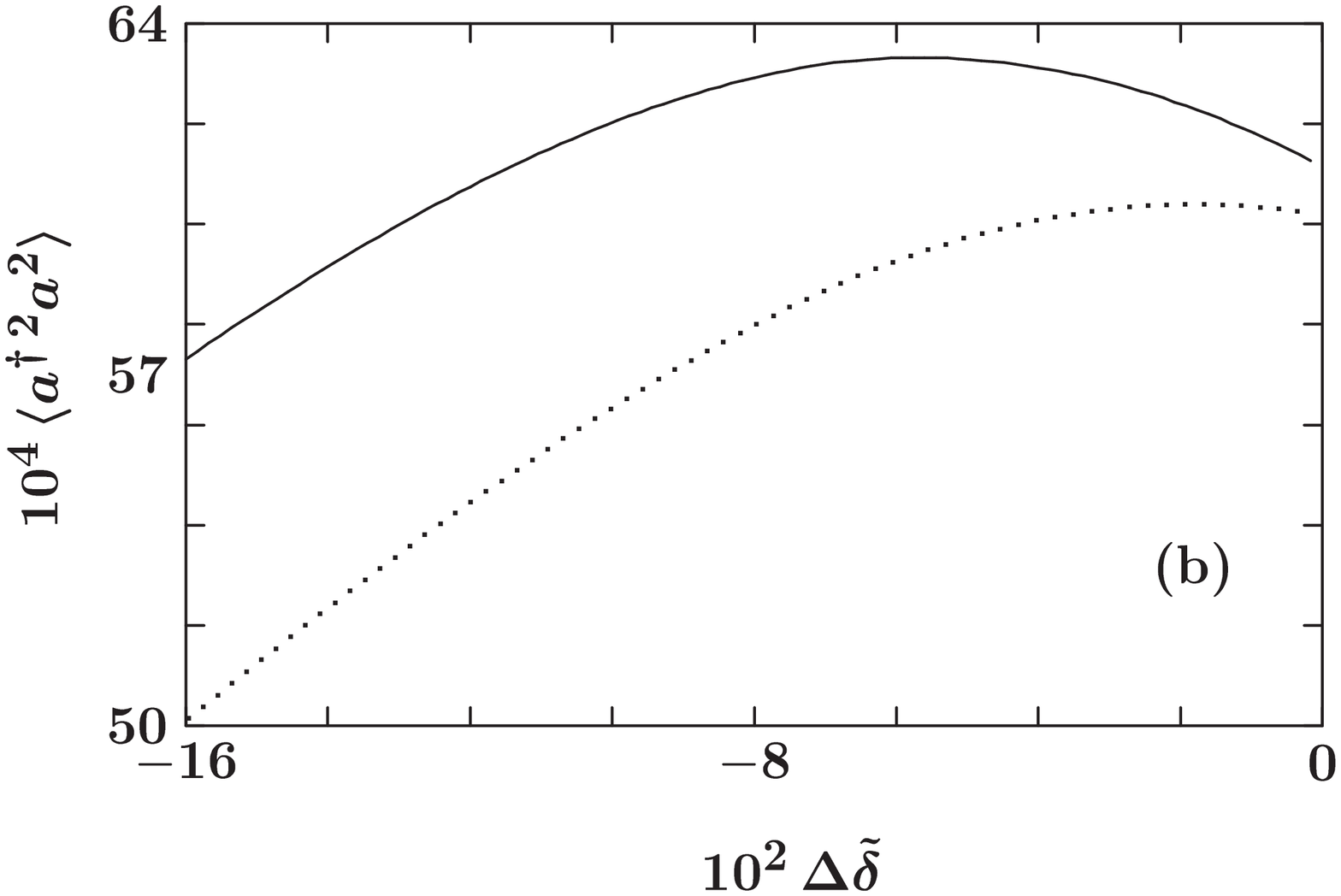}}}
\end{picture}
\caption{The 2PCR vs peak shift from~$\tilde\delta=\sqrt2+1$ for
	the inhomogeneously broadened system with
	${\cal E}_1/\kappa=1/\sqrt{2}$,
	${\cal E}_2/\kappa=\sqrt{2}$, $g_f/\kappa=9$, $\gamma/\kappa=2$. 
	Two cases are shown:
	(a) with (solid) and without (dots) the imposition of
	condition~(\ref{pathwayrestriction}); and (b) with
	the imposition of 
	condition~(\ref{pathwayrestriction}) and the
	homogeneous linewidth (and jump terms) set to
	zero for the~$\vert 1\rangle_-$ level of the JC ladder
	(solid), and with the additional imposition that
	~$\langle 0\vert\Upsilon({\cal E}_2\exp(-i\delta t))\vert 1 \rangle_+$
	and its complex conjugate are set to zero (dots).}
\label{fig:fig4}
\end{figure}

Thus far our attention has been restricted to the $P(g)=\delta(g-g_f)$
case.  This case is a significant contributor to the peak shift for the 
inhomogeneously--broadened peak shift obtained in the full simulation,
accounting for atomic position variability.  The simulation of the 2PCR 
peak near $\tilde{\delta}=\sqrt{2}+1$ is depicted in Fig.~\ref{fig:fig4}(a) 
as a solid line.
This solid line evidently has a peak at $\Delta\tilde{\delta}=-0.141$.  The
above reasoning, for the $P(g)=\delta(g-g_f)$ case, suggests that the
competing pathway $\vert 1 \rangle_- \longleftrightarrow \vert 2\rangle_-$ is
partially responsible for this peak shift. Therefore, we repeat the simulation,
but with the restriction~(\ref{pathwayrestriction}).  The result of this 
simulation is also depicted in Fig.~\ref{fig:fig4}(a) 
as a dashed line.  The new peak position is
located approximately at~$\Delta\tilde{\delta}=-0.094$.  This peak position
can be shifted slightly by eliminating other pathways.
By fixing all transition terms of the type~(\ref{pathwayrestriction})
to be zero {\em except} 
for~$\langle 0\vert\Upsilon({\cal E}_1)\vert 1 \rangle_-$,
$_-\langle 1\vert\Upsilon({\cal E}_2 \exp(-i\delta t))\vert 2 \rangle_+$,
and their complex conjugates, the resultant peak is centered 
at~$\Delta\tilde{\delta}=-0.091$. These additional pathways, which 
were ignored in obtaining~$\Delta\tilde\delta=-0.094$, 
do not contribute much to the overall peak shift, 
and we do not depict this specific peak shift.

To account for the remainder of the peak shift observed in 
Fig.~\ref{fig:fig4}(a), we determine other pathways to an enhanced 2PCR, 
which arise in the full inhomogeneously broadened case but not in the 
$g=g_f$ case.
We now consider the effect of the homogeneous linewidth at the 
level~$\vert 1 \rangle_-$ state. To incorporate this effect, we solve 
the dynamics for the effective Hamiltonian~(\ref{eq:Heff}) plus the driving 
term~(\ref{Liouvillean:D}) but do not include the jump 
term~(\ref{Liouvillean:J}).  The 2PCR, which eliminates the pathway
$\vert 1 \rangle_- \longleftrightarrow \vert 2\rangle_-$
and the homogeneous linewidth at~$\vert 1 \rangle_-$ 
is depicted in Fig.~\ref{fig:fig4}(b).  
We observe that the resultant peak shift is
now $\Delta\tilde{\delta}=-0.057$.  The consideration of the homogeneous
linewidth does matter in accounting for the peak shift.

To understand the peak shift due to the homogeneous
linewidth at $\vert 1 \rangle_-$,
we consider the following.  The pathway $\vert 0 \rangle
\longleftrightarrow \vert 1
\rangle_- \longleftrightarrow \vert 2 \rangle_+$ is relatively 
unaffected by the 
homogeneous linewidth as it is a resonant transition
at $\Delta\tilde{\delta}=0$, but the homogeneous linewidth does enable
off--resonant excitation to $\vert 2 \rangle_-$, even after the restriction
(\ref{pathwayrestriction}) has been applied.
The homogeneous linewidth of the $\vert 1 \rangle_-$ level enables
two--photons transitions to take place for a range of coupling strength~$g$, 
via the sequence of an $\omega_1$ photon and then
an $\omega_2$ photon. By eliminating this homogeneous linewidth, the
$\omega_1$ photons really must satisfy the condition 
$\omega-\omega_1=g_f$ in order to 
off--resonantly excite to the $\vert 2 \rangle_-$ level.  The homogeneous
linewidth is responsible for permitting competing pathways from
$\vert 0 \rangle\longleftrightarrow\vert 1 \rangle_-
\longleftrightarrow\vert 2 \rangle_-$
for a range of $g$ values, and eliminating this homogeneous linewidth
eliminates a whole class of competing pathways for the inhomogeneously 
broadened (general $P(g)$) system.

The final significant contribution to the peak shift of the 2PCR curve
arises for small~$g$.  The low coupling strength limit is important
because the distribution~$P(g)$ is so heavily weighted in favor of
low~$g$.  When the coupling strength is small, the splitting between
the first couplet states~$\vert 1 \rangle_{\pm}$ becomes correspondingly
small.  Consequently one would expect a competing pathway to arise via
excitation to $\vert 1 \rangle_+$.  We eliminate the contribution due to the
pathway $\vert 0 \rangle \longleftrightarrow \vert 1 \rangle_+$.  
Incorporating this restriction with the other restrictions previously \
mentioned (eliminating the $\vert 1 \rangle_- 
\longleftrightarrow \vert 2 \rangle_-$ pathway and the homogeneous linewidth
at $\vert 1 \rangle_-$) leads to the dotted curve in Fig.~\ref{fig:fig4}(b).  
The resultant
peak shift is just $\Delta\tilde{\delta}=-0.018$.

The three effects considered
above account for 87\% of the peak shift.  Better agreement can be obtained
by reducing homogeneous linewidth on the $\vert 2 \rangle_-$ level
or by eliminating transitions to the competing state~$\vert 2 \rangle_-$
altogether.  However, the agreement demonstrated here exhibits excellent
agreement with the full simulated 2PCR peak at $\tilde{\delta}=\sqrt{2}+1$.
There is no doubt that this shift is primarily due to
competing pathways in the JC ladder.

In accounting for the peak shift due to competing pathways, we have included
restrictions of coherent transitions and thereby accounted for the majority
of the peak shift.  There is also competition with incoherent pathways,
which are accounted for by the quantum jump terms.  The quantum jumps
correspond to the transfer of quanta between levels of the JC ladder,
but the contribution of these jumps to peak shifts is not large for the
bichromatically--driven JC system.  Never the less, the jump terms are 
responsible for part of the peak shift. This competition with incoherent
pathways is quite pronounced for the monochromatically--driven three--level
system discussed in the next section.

\section{The Monochromatically--driven three--level $\vee$--system}
\label{sec:MDTLVS}

The shift in spectral peaks should not be unique to
a bichromatically--driven JC system.
A simple case to consider is the monochromatically--driven three--level
$\vee$--system (the JC ladder in the linear regime, cf Fig.~\ref{fig:ladder}),
which is of interest for studying normal--mode (or vacuum Rabi) 
splitting~\cite{Rai89,Zhu90,Tho92}.

We consider the three--level truncation of the JC ladder 
to~$\left\{ \vert 0\rangle, \vert 1\rangle_{\pm} \right\}$ and
a monochromatic driving field which is resonant with
the~$\vert 0\rangle\longleftrightarrow\vert 1\rangle_+$ transition 
($\omega_1=\omega+g_f$) where~$\omega_1$ is allowed to vary.
In contrast to the previous section, enhanced one--photon count rates
identify resonances and correspond to the standard method of spectroscopy,
as opposed to photon {\em coincidence} spectroscopy.  
The normalized frequency of this driving 
field is~$\tilde\delta=-\delta/g_f$ for~$\delta=\omega-\omega_1$.
We expect a peak at~$\tilde\delta=1$, for which resonant
excitation occurs
if we ignore competition between excitation pathways. 

\begin{figure}
\begin{picture}(200,180)(60,10)
\font\gnuplot=cmr10 at 10pt
\gnuplot
\rotatebox{90}{\resizebox{250pt}{330pt}{\includegraphics{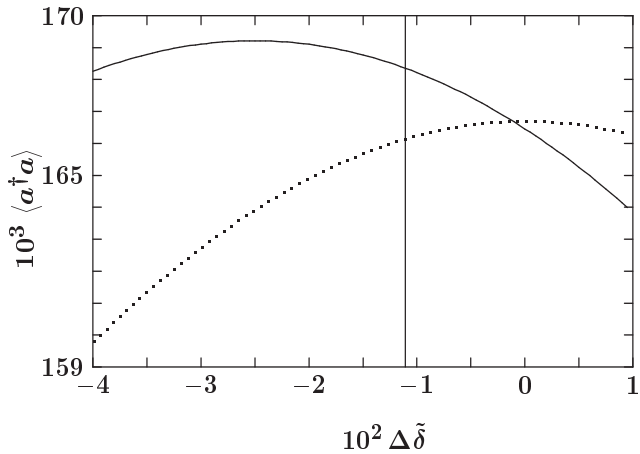}}}
\end{picture}
\caption{One--photon count rate (1PCR) of a three--level
	system vs the shift in normalized 
	scanning 
	frequency of the peak at~$\tilde\delta=1$ 
	for~$g/\kappa=9$, 
	${\cal E}_1/\kappa=\sqrt{2}$ and $\gamma/\kappa=2$
	depicted as the solid line. 
	The dotted line corresponds to the 
	case~$\langle 0\vert \Upsilon({\cal E}_1\exp(-i\delta t))
	\vert 1\rangle_-=0=_-\!\!\langle 1\vert \Upsilon({\cal E}_1
	\exp(-i\delta t))\vert 0\rangle$. The vertical line
	corresponds to~$\Delta\tilde\delta$ in Eq.~(\ref{eq:peakshift}).}
\label{fig:shiftJC3}
\end{figure}

In Fig.~\ref{fig:shiftJC3} we see that,
for~$\Delta\tilde\delta=\tilde\delta-1$, the peak is shifted to the left.
Competition due to off--resonant excitation is established by
setting certain matrix elements of the driving term in the
master equation to zero, by analogy with the 2PCR considered above.
That is, for~$H_{\rm eff}(g,{\cal E}_1)$ in Eq.~(\ref{eq:Heff}), we 
impose
\begin{equation}
\label{eq:competitionterm}
\langle 0\vert\Upsilon({\cal E}_1 \exp(-i\delta t)) 
\vert 1 \rangle_-=0= _-\!\!\langle 1\vert\Upsilon({\cal E}_1 \exp(-i\delta t)) 
\vert 0\rangle
\end{equation}
in the simulation. We observe that this condition
produces a plot of~$\langle a^{\dag}a\rangle$ that is exactly
centered at~$\Delta\tilde\delta=0$. Therefore, the competition with the
off--resonant excitation pathway~$\vert 0\rangle\longleftrightarrow 
\vert 1\rangle_-$
is entirely responsible for the peak shift of~$\langle a^{\dag}a\rangle$ 
vs~$\tilde\delta$.

Another perspective for understanding the spectral peak shift for the 
driven $\vee$--system is obtained directly from the effective
Hamiltonian of Eq.~(\ref{eq:Heff}) solved in the linear regime.
We define for the pseudo--pure 
state~$\psi$,~$C_0\equiv\langle 0\vert\psi \rangle$ 
and~$C_{1\pm}\equiv _{\pm}\langle 1\vert\psi \rangle$. The resulting
set of differential equations for these dressed--state amplitudes
$C_0$, $C_{1-}$ and $C_{1+}$ is given by
\begin{eqnarray}
\label{eq:3leveldiff1}
\dot{C}_0&=&-\frac{{\cal E}_1}{\sqrt{2}}(C_{1-}+C_{1+}), \\
\label{eq:3leveldiff2}
\dot{C}_{1-}&=&\frac{{\cal E}_1}{\sqrt{2}} C_0-
\left( i(\delta - g) +\frac{1}{2}\left(\kappa
+\frac{\gamma}{2}\right)\right)C_{1-} \nonumber \\
& &+\frac{1}{2}\bigg(\kappa-\frac{\gamma}{2}
\bigg)C_{1+} \\
\label{eq:3leveldiff3}
\dot{C}_{1+}&=& \frac{{\cal E}_1}{\sqrt{2}}C_0
-\left( i(\delta + g) 
+\frac{1}{2}\left(\kappa
+\frac{\gamma}{2}\right)\right)C_{1+} \nonumber \\
& &+\frac{1}{2}\bigg(\kappa-\frac{\gamma}{2}\bigg)C_{1-}.
\end{eqnarray}
Taking the Laplace
transform in the long time limit, a minimum and two maxima
occur in the PCR for 
\begin{eqnarray}
\label{1}
\tilde\delta &=& \tilde\delta_0 \equiv 0 \; \mbox{(minimum)}~,\\ 
\label{2}
\tilde\delta &=& \tilde\delta_\pm
	\equiv -\sqrt{1 \pm \frac{2+{\cal E}^2}{2g^2}} \; \mbox{(maxima)},
\end{eqnarray} 
for the decay rate~$\gamma/\kappa=2$. 
To identify competition between pathways, we impose the condition
Eq.~(\ref{eq:competitionterm}). This condition modifies
Eq.~(\ref{eq:3leveldiff1}) by replacing the~$C_{1-}$ term 
by zero and modifies Eq.~(\ref{eq:3leveldiff2}) 
by replacing~$C_0$
by zero. 
The peak at~$\tilde\delta_-$
thus vanishes, as there
is no driving to the~$\vert 1\rangle_-$ state. The 
peak 
at~$\tilde\delta_+$
is shifted 
to~$\tilde\delta=1$ (where the peak would be expected 
initially without any competition between excitation pathways). 
The formula for the
competition--induced peak shift for the three--level JC system is thus given by
\begin{equation}
\label{eq:peakshift}
\Delta\tilde\delta=\tilde\delta_+-1=\sqrt{1-\frac{2+{\cal E}^2}{2g^2}}-1.
\end{equation}

The peak shift~$\Delta\tilde\delta$ presented in~Eq.~(\ref{eq:peakshift})
accounts partially for the total peak shift presented in 
Fig.~(\ref{fig:shiftJC3}). The discrepancy is due solely to the effect of 
quantum jump terms, which have been ignored in Eqs.~(\ref{1})--(\ref{2}).
The necessity of including quantum jump terms arises because 
dissipation terms cannot be dissociated from fluctuations in 
a proper analysis.
The quantum jump terms are responsible for incoherent transitions of quanta
from one level to another of the three--level system.
We observe, in Eqs.~(\ref{1})--(\ref{2}), that incoherent transitions are
not included in the mathematics.  Specifically, we can see that diagonal
matrix elements $\rho_{00}=C_0^*C_0$ and $\rho_{11}^{\pm\pm}=C_{1\pm}^*
C_{1\pm}$ are not coupled together.  This coupling of diagonal elements of the
density matrix arises via quantum jump terms.  These quantum jump terms are 
responsible for the discrepancy between the predicted peak shift given
by Eq.~(\ref{eq:peakshift}) and the observed peak center of the solid
locus in Fig.~\ref{fig:shiftJC3}.  Whereas, for the bichromatically--driven
JC system, the contribution of incoherent pathways to the net peak shift
is small, the contribution of incoherent pathways to the peak shift for
the monochromatically--driven three--level $\vee$ system is relatively large
because there are fewer coherent pathways to producing a peak shift as 
compared to the bichromatically--driven JC system considered in the previous
section.

The advantage of using Eqs.~(\ref{1})--(\ref{2}), to confirm that
the peak shift is indeed partially due to competition between 
coherent excitation pathways, is that this method is entirely analytical.
The analytical method provides a simpler conceptual framework to 
understand the role of both coherent and incoherent excitation pathways.
In this analytical method, we obtain solutions via the Laplace
transform method rather than relying on numerical solutions to the
full master equation. Numerical
methods are necessary, though, to fully account for spectral peak shifts.

\section{Conclusion}
\label{sec:Conclusion}

The analysis presented here provides convincing evidence that the
shifts in two--photon coincidence spectral peaks, from expected
locations suggested by the JC ladder of Fig.~\ref{fig:ladder}, are due
overwhelmingly to competition between excitation pathways.

We have analyzed in detail the two--photon spectral peaks, outside the
domain where background subtraction is essential for resolution. 
The shift of the spectral peak has been shown to be
a consequence of competition between excitation pathways. This 
demonstration relies on numerical methods whereby the competing
pathway have been artificially eliminated by imposing the condition
that certain matrix elements are zero in the master equation.

The key contribution to the peak shift were identified to be 
competition with the~$\vert 0\rangle \longleftrightarrow \vert 1\rangle_-
\longleftrightarrow \vert 2\rangle_-$ 
pathway for~$g\approx g_f$,~$\vert 0\rangle 
\longleftrightarrow \vert 1\rangle_+\longleftrightarrow \vert 2\rangle_{\pm}$ 
for~$g\rightarrow 0$ and the homogeneous linewidth about~$\vert 1\rangle_-$
which allowed off--resonant excitation to~$\vert 2\rangle_{\pm}$.
Eliminating these three effects reduces the two--photon coincidence
peak shift by~$87\%$.

These shifts in spectral peaks are not restricted to 2PCS. To
verify this, we considered the shifts of spectral peaks for 
standard (one--photon) spectroscopy and a three--level 
monochromatically--driven $\vee$--system. Such a system corresponds to
the driven JC system in the linear regime, which is of interest in
the study of normal--mode (vacuum Rabi splitting). We also
observe the shift of spectral peaks in our simulations. By ignoring the
jump terms, the equations become easily solvable and 
we verify that a significant part of the peak shift for the $\vee$--system are
indeed due to competition between excitation pathways.  The analytical
method enables us to clearly see the distinction between coherent and 
incoherent pathways as competition--induced peak shifts.  The contribution
of quantum jump terms to the net peak shift is significant for the
monochromatically--driven $\vee$ system, whereas the effect is much
smaller for the bichromatically--driven JC system.  The combination of 
competition with coherent and incoherent pathways fully accounts for the
peak shift of the monochromatically--driven three--level $\vee$ system
analyzed in Sec.~\ref{sec:MDTLVS}.

These results clarify the expected peak locations in 2PCS and should
also apply to the more general Multiphoton Coincidence 
Spectroscopy~\cite{Hor99}. As photon coincidence spectroscopy must be 
accurate to verify the signature of entanglement, these peak shifts
must be fully accounted for and properly understood. The analysis
presented here does indeed explain the peak shifts clearly and 
quantitatively.

\acknowledgments
We acknowledge useful discussions with H.~J.\ Carmichael, Z.\ Ficek, 
K.--P.\ Marzlin and Weiping Zhang.  This research has been supported 
by Australian Research Council Large and Small Grants and by an Australian 
Research Council International Research Exchange Scheme Grant.

\end{document}